\documentclass[12pt]{article}

\usepackage{german}
\usepackage{a4wide}

\usepackage{amsmath,amssymb,amsfonts} 

\usepackage{mathrsfs,latexsym}

\numberwithin{equation}{section}

\begin{document}

%

\title{Zeitreisen: Gravitation trifft Quantenphysik}
\author{Rainer Verch \\[6pt]
{\small Institut f\"ur Theoretische Physik,
Universit\"at Leipzig,
04009 Leipzig, Germany}}
\maketitle
\begin{abstract} \noindent
Es wird eine Einf\"uhrung gegeben in die Diskussion, ob es m\"oglich ist, mit Hilfe von Materie,
die typisch quantenphysikalische Eigenschaften aufweist, Raumzeit-Geometrien herzustellen, in denen
Zeitreisen m\"oglich sind. Die m\"oglichen Szenarien werden im Rahmen der Quantenfeldtheorie
und der allgemeinen Relativit\"atstheorie betrachtet. Es werden in der Literatur etablierte
Ergebnisse dargestellt, wonach die M\"oglichkeit f\"ur Zeitreisen im wesentlichen ausgeschlossen ist.
\\[6pt]
{\it An introduction is given to discussions on the possiblity of fabricating spacetime geometries allowing
time-travel scenarios with the help of matter possessing typically quantum features. Those scenarios are
considered in the framework of quantum field theory and general relativity. Results established in the 
literature excluding the possibility of time-travel are presented.}
\end{abstract}

\section{Einf\"uhrung: Gravitation und Quantenphysik}
Zeitreisen, also die M\"oglichkeit, nicht nur im Raum von einem Ort zum anderen zu reisen,
sonderen von einem Zeitpunkt zu einem anderen, etwa von der Gegenwart in die Vergangenheit oder
in die Zukunft, bieten faszinierende Perspektiven. Tats\"achlich scheint es einem, wir h\"atten uns
schon daran gew\"ohnt, Zeitreisen als ein handlungstreibendes Element in der Belletristik und in
Unterhaltungsfilmen zu akzeptieren; in vielen Filmproduktionen der letzten beiden Jahrzehnte kommen
Zeitreisen so h\"aufig vor, dass man fast glaubt, die geh\"orten zur zeitgem\"a{\ss}en Kinounterhaltung einfach
dazu --- und, wohlgemerkt, nicht nur auf das Science-Fiction Genre beschr\"ankt.

Nat\"urlich wissen wir, dass sich Zeitreisen nicht gut mit unserer Alltagserfahrung in \"Ubereinstimmung
bringen lassen, und es gibt bisher kein seri\"oses, best\"atigtes und unumstrittenes wissenschaftliches Experiment, das 
nachgewiesen h\"atte, physikalische Systeme oder Signale in die Zukunft oder die Vergangenheit zu versenden zu k\"onnen.
Es gibt jedoch durchaus eine wissenschaftliche Diskussion \"uber die M\"oglichkeit --- oder Unm\"oglichkeit ---
von Zeitreisen. Denn der Umstand, dass Zeitreisen bisher nicht zu unserem Erfahrungsbereich geh\"oren,
schliesst ja ihre prinzipielle Durchf\"uhrbarkeit nicht unbedingt aus; es k\"onnte sein, dass dazu extreme
Bedingungen hergestellt werden m\"ussen, die \"ublicherweise nicht vorkommen, aber mit gen\"ugend fortgeschrittener
Technologie erreichbar w\"aren. Um solche grunds\"atzlichen \"Uberlegungen, oder jedenfalls einen Teilaspekt
solcher \"Uberlegungen, soll es hier gehen. Dabei sollte betont werden: Hier wird nicht der Versuch
unternommen, die durchaus schon recht umfangreiche Literatur, die dieser Fragestellung gewidmet ist, auch nur
ansatzweise repr\"asentativ darzustellen. Es soll auch nicht auf die einigerma{\ss}en bekannten ``Paradoxa'' eingegangen
werden, die sich im Zusammenhang mit der Frage nach der Durchf\"uhrbarkeit von Zeitreisen ergeben. Einen recht
guten ersten \"Uberblick zu dem Thema generell bieten die Artikel der Stanford Encyclopedia of Philosophy
\cite{SEPH-TiMach,SEPH-TiTrv}. Eine sehr lesenwerte und bemerkenswert
umfassend recherchierte Darstellung vieler Aspekte des Themas bietet das vorz\"ugliche Buch von Paul Nahin \cite{Nahin}.
Eine mehr wissenschaftliche Diskussion, in der Fachsprache der daf\"ur relevanten Teilgebiete der
Theoretischen Physik, wird in einem aktuellen Artikel der Wissenschaftsphilosophen Earman, Smeenk und W\"uthrich
gegeben \cite{EarSmeWut}.

Hier geht es um die Frage nach der M\"oglichkeit von Zeitreisen im Rahmen der bestehenden Erkenntnisse
der (Theoretischen) Physik, wobei zwei Hauptrichtungen besonders betont werden: Gravitation und Quantenphysik.
Genauer sollte spezifiziert werden: Die Beschreibung der Gravitation und der Struktur von Raum und Zeit im
Sinne der Einsteinschen allgemeinen Relativit\"atstheorie, und der Quantenphysik im Sinne der Quantenfeldtheorie.

Im Standardkanon der Ausbildung eines Physikers sind die allgemeine Relativit\"atstheorie und die Quantenphysik, oder
genauer die Quantenfeldtheorie, Theorien f\"ur sehr verschiedene Ph\"anomenbereiche. Die allgemeine Relativit\"atstheorie
ist die Theorie der Gravition; als eine der fundamentalen Wechselwirkungen betrachtet, ist die Gravitation die
Schw\"achste aller bekannten, aber daf\"ur diejenige, die bei gro{\ss}en Masseansammlungen und \"uber gro{\ss}e
Entfernungen wirksam ist; tats\"achlich ist die Gravitation die f\"ur das Universum strukturbildende Wechselwirkung.
Dagegen werden Prozesse der Materie auf den kleinsten Skalen durch die Quantenphysik beschrieben. Hier geht es also
um die Physik von Atomen und Elementarteilchen bis hinunter zu den kleinsten bekannten L\"angen- und Zeitskalen, unter
Beteiligung aller bekannten Wechselwirkungen au{\ss}er der Gravitation (elektroschwache und starke Wechselwirkung).
In der Elementarteilchenphysik beschreibt die Quantenfeldtheorie auch die Umwandlung verschiedener Teilchensorten
bei Prozessen mit sehr hohem Energieaustausch, wie sie beispielsweise in Elementarteilchenbeschleunigern herbeigef\"uhrt
und analysisert werden.

Es stellt sich die Frage, ob es Situationen gibt, in denen sowohl die Gravitation wie auch die Quantenphysik
ganz wesentlich in die Beschreibung eingehen. Dies ist zu erwarten bei Prozessen mit sehr hoher Energiedichte, die
extreme Gravitationsgradienten erzeugt. Zu solchen extremen Situation z\"ahlen der Kollaps eines Sterns zu einem schwarzen
Loch (insbesondere der --- v\"ollig unbekannte --- Materiezustand im Inneren eines schwarzen Lochs) oder der 
``Anfangszustand'' des Universums,  wo nach den g\"angigen Theorien der Kosmologie die Materie und
die Geometrie der Raumzeit singul\"are Eigenschaften haben m\"ussten, die im Rahmen der Quantenfeldtheorie und
der Einsteinschen allgemeinen Relativit\"atstheorie bisher nicht beschrieben werden k\"onnen. Die allgemeine Erwartung ist,
dass es eine Theorie der Quantengravitation geben sollte, die auch die Struktur von Raum und Zeit als ``quantisiert''
beschreibt. Was genau das bedeutet ist nicht wirklich klar, denn
ein wesentlicher Zug der quantenphysikalischen Beschreibung von Prozessen sind Unsch\"arferelationen zwischen
bestimmten Me{\ss}gr\"o{\ss}en (Observablen) der Theorie, die unter anderem Grenzen einer deterministischen Beschreibung
von Einzelereignissen bedeuten. Eine quantisierte Struktur von Raum und Zeit h\"atte m\"oglicherweise zur Folge,
dass die zeitliche Anordnung von physikalischen Vorg\"angen nicht mehr ohne weiteres gegeben ist. In einem solchen Rahmen
deutet sich eine M\"oglichkeit f\"ur Zeitreisen durchaus an. Allerdings gibt es bisher keine vollst\"andige Theorie
der Quantengravitation, sondern nur einige Ans\"atze dazu, die weiter erforscht werden. Die prominentesten Ans\"atze
sind: ``Loop Quantum Gravity'' \cite{Thiemann}, Stringtheorie \cite{Zwiebach} und nicht-kommutative Geometrie \cite{Connes}
--- die hier genannten Literaturhinweise bieten einen Einstieg in die jeweiligen Gebiete.

Hier wird der Standpunkt eingenommen, dass die Ans\"atze f\"ur eine Theorie der Quantengravitation noch nicht
gen\"ugend weit entwickelt sind, um als Grundlage f\"ur eine Diskussion \"uber die M\"oglichkeit von Zeitreisen
zu dienen. Vielmehr soll hier vom einem verh\"altnism\"a{\ss}ig konservativen Ansatzpunkt ausgegangen werden:
Der Quantenfeldtheorie im Zusammenhang mit der Einsteinschen Gravitationstheorie als klassischer, im Sinne von
``nicht quantisierter'', Theorie. Diese Theorie wird allgemein auch mit dem Begriff ``Quantenfeldtheorie in
gekr\"ummter Raumzeit'' bezeichnet. Einige ihrer wesentlichen Bestandteile werden sogleich vorgestellt werden;
zuvor sei jedoch gesagt, dass diese Theorie den Hawking-Effekt und den Unruh-Effekt beinhaltet, sowie die
Quantisierung der Fluktuationen in der fr\"uhen inflation\"aren Epoche der Kosmologie, die wiederum als
urs\"achlich f\"ur die Temperaturschwankungen in der kosmischen Hintergrundstrahlung angesehen werden. W\"ahrend
Quantenfeldtheorie in gekr\"ummter Raumzeit vielleicht nicht das leisten kann, was man von einer Theorie der
Quantengravitation erwarten w\"urde, so liefert sie doch die Grundlage zur quantitativen Beschreibung einiger
Effekte, von denen man denkt, dass sie auch in der Quantengravitation relevant sind. Ein recht zug\"anglicher
Einstieg in die Quantenfeldtheorie in gekr\"ummter Raumzeit findet sich in einem Buch von Robert Wald
\cite{WaldQFT}, aktuelle \"Ubersichtsartikel sind z.B. \cite{BDH,Ver-Reb}.

\section{Allgemeine Relativit\"atstheorie, Gravitation und Zeitreisen}

An dieser Stelle muss nun zun\"achst die Einsteinsche allgemeine Relativit\"atstheorie kurz vorgestellt werden, was
zwangsl\"aufig die Darstellung sehr viel technischer macht als bisher. Sehr gute Lehrb\"ucher \"uber die
allgemeine Relativit\"atstheorie sind z.B. \cite{WaldGR,Goenner}. 

Eine Grundannahme der allgemeinen Relativit\"atstheorie besteht darin, dass alle Ereignisse, d.h.\ alle physikalischen
Vorg\"ange, unabh\"angig davon, ob sie m\"oglich oder faktisch sind, sich in einer Menge zusammenfassen lassen, die die Struktur
einer vierdimensionalen differenzierbaren Mannigfaltigkeit (im Sinne der Differentialgeometrie) hat. Die vier
Dimensionen bedeuten eine reelle Zeitkoordinate und 3 reelle r\"aumliche Koordinaten, mit denen die Ereignisse
gekennzeichnet werden k\"onnen. Jedes Ereignis steht also f\"ur einen Zeitpunkt und einen Punkt im Raum. Die Menge
aller Ereignisse nennt man auch Raum-Zeit-Kontinuum oder einfach Raumzeit. Wir bezeichnen die Raumzeit mit $M$.
Eine weitere zentrale Annahme ist, dass auf der Raumzeit $M$ eine Metrik $g_{ab}$ mit Lorentzscher Signatur
definiert ist. Die Metrik bestimmt die Raumzeit-Geometrie: Materielle Objekte bewegen sich entlang zeitartiger
Geod\"aten, Lichtsignale entlang lichtartiger Geod\"aten. Ferner werden die Effekte von Gravitation bestimmt
durch durch die Kr\"ummungsgr\"o{\ss}en, die der Metrik zugeordnet sind, wie dem Riemannschen Kr\"ummungstensor, dem
Riccitensor und der Skalarkr\"ummung. Die Kr\"ummungsgr\"o{\ss}en bestimmen, ob ein Schwarm zeitartiger Geod\"aten
voneinander wegdriftet oder sich aufeinander zubewegt. 

Der ganz wesentliche Grundsatz der Einsteinschen allgemeinen Relativit\"atstheorie ist dabei, dass die Raumzeit-Geometrie
nicht fest vorgegeben ist (so wie in der speziellen Relativit\"atstheorie und in der Newtonsch-Galileischen Theorie
von Raum und Zeit), sondern dynamisch bestimmt wird durch die in der Raumzeit vorhandene Energie und Materie.
Dies wird ausgedr\"uckt durch die Einsteinschen Feldgleichungen der Gravitation:
$$
    R_{ab} - \frac{1}{2} g_{ab} R + \Lambda g_{ab} = 8\pi G T_{ab} \,.
    $$
Auf der linken Seite der Gleichung stehen die Raumzeitmetrik $g_{ab}$ und ihre Kr\"ummungsgr\"o{\ss}en, wie der
Ricci-Tensor $R_{ab}$ und die Skalarkr\"ummung $R$. Die Gr\"o{\ss}e
$\Lambda$ ist die sogenannte kosmologische Konstante. Sie kann f\"ur die meisten Betrachtungen gleich null gesetzt werden,
oder aber es kann anstelle des Terms $\Lambda g_{ab}$  auf der linken Seite ein Term
$-\Lambda g_{ab}$ auf die rechte Seite geschrieben werden --- was allerdings die Interpretation des Terms
ab\"andert. Die Diskussion zu dieser Auffassung ist durchaus vielschichtig und hat, besonders f\"ur die
Kosmologie, durchaus Konsequenzen \cite{PeebRh}, aber wir werden diese Angelegenheit nicht ber\"uhren.
Auf der rechten Seite steht der Energie-Impuls-Tensor der
Materie und Energie, die in der Raumzeit vorhanden oder verteilt ist. Dabei ist $G$ die Newtonsche Gravitationskonstante.
Der Energie-Impuls-Tensor beschreibt den Energieinhalt aller Materie- und Energieformen im Sinne der Hydrodynamik.
Tats\"achlich kommt durch die Einsteinschen Feldgleichungen ein dynamisches Wechselspiel zwischen Raumzeit-Geometrie
und dem Energieinhalt der Materie in der Raumzeit zum Ausdruck: Die Anwesenheit von Energie und Materie erzwingt die
Kr\"ummung der Metrik. Andererseits erzwingt die Kr\"ummung gravitative Kr\"afte, also die Bewegung der Materie relativ
zueinander. Tats\"achlich besagt die Erfahrung, dass Gravitation immer anziehend wirkt, und daher sich die 
Materie immer aufeinander zubewegt (sofern sie gen\"ugend konzentriert ist). Dieser attraktive Charakter der Gravitation
geht einher mit Positivit\"at der Energiedichte. Dies bedeutet, dass jeder Beobachter in der Raumzeit eine positive
Energiedichte misst. Ein Beobachter wird als eine zeitartige Kurve $\gamma$ in der Raumzeit dargestellt, das heisst durch
seine Weltlinie --- wobei ``Beobachter'' hier synomym f\"ur ``Me{\ss}ger\"at'' steht. Der relativistische Geschwindigkeitsvektor
der Kurve ist $u^a = (d\gamma/dt)^a$, wobei $t$ die Eigenzeit des Beobachters entlang seiner Weltlinie ist.
Die Bedingung $g_{ab}u^a u^b > 0$ bedeutet, dass die Kurve zeitartig ist, also sich mit einer Geschwindigkeit geringer als 
Lichtgeschwindigkeit (in dieser Weise festgelegt durch die Raumzeit-Metrik $g_{ab}$) in der Raumzeit bewegt. Dann ist
$$ \varrho[u] = T_{ab}u^a u^b $$
die Energiedichte, die der Beobachter an seinem momentanen Aufenthaltsort misst, wenn der Energie-Impuls-Tensor den
gesamten umgebenden Energie- und Materieinhalt beschreibt. Die Bedingung positiver Energiedichte lautet dann
$$ \varrho[u] \ge 0 $$
f\"ur jeden Beobachter, und an jedem Raum- und Zeitpunkt in der Raumzeit.

Die Einsteinschen Feldgleichungen sind genau genommen als ein Anfangswertproblem zu formulieren. Das heisst, es
werden zu einer gewissen Anfangszeit die Metrik $g_{ab}$ und Kr\"ummungsgr\"o{\ss}en \"uberall im Raum
vorgegeben, und dazu Anfangsdaten der Energie-Materieverteilung \"uberall im Raum. Zu\"atzlich gibt es Bewegungsgleichungen
f\"ur die Energie-Materie-Freiheitsgrade. Werden diese zusammen mit den Einsteinschen Feldgleichungen gel\"ost, so
ergibt sich die Raumzeit-Geometrie zusammen mit der Energie-Materie-Verteilung in der Zukunft und der Vergangenheit
der gew\"ahlten Anfangszeit. Ein Argument von Hawking besagt dann, dass die Bedingung positiver Energiedichte das 
Auftreten von Zeitmaschinen (oder Zeitreisen) unter recht allgemeinen Voraussetzungen
verhindert \cite{Haw92}. 
Hier tritt erstmals der Begriff einer Zeitmaschine, hier synonym gebraucht zum Begriff Zeitreise, auf, und
hierf\"ur muss nun eine Definition gegeben werden. Das ist nicht einfach, und tats\"achlich ist nicht
unumstritten, was unter dem Begriff einer Zeitmaschine oder Zeitreisen im Kontext der allgemeinen 
Relativit\"atstheorie verstanden werden soll. Einigerma{\ss}en unumstritten ist jedoch die Vorstellung, dass
Zeitreisen das Auftreten von geschlossenen zeitartigen Kurven beinhaltet.
Eine zeitartige Kurve in der Raumzeit ist, wie erw\"ahnt, eine glatte Abbildung
$\gamma : I \to M$, $t \mapsto \gamma(t)$ von einem reellen Intervall $I$ in die Raumzeit-Mannigfaltigkeit $M$, 
mit $g_{ab}u^a u^b > 0$ f\"ur jeden Tangentialvektor $u^a$ entlang der Kurve. Sie
repr\"asentiert ganz allgemein die Weltlinie eines (idealisiert ausdehnungslosen) materiellen Objekts --- eines 
Beobachters, eines Messger\"ats, oder eines (Elementar-)Teilchens; aber je nach Kontext auch eines Himmelk\"orpers oder
einer Galaxie. Die zeitartige Kurve heisst geschlossen, wenn es unterschiedliche Parameterzeiten $t_1 \ne t_2$ in $I$ gibt
so, dass $\gamma(t_1) = \gamma(t_2)$ gilt, so dass also der Beobachter sich zu verschiedenen Zeiten seiner mitgef\"uhrten
Uhr (deren Zeigerstellungen durch den Parameter $t$ angegeben werden) an demselben Raum- und Zeitpunkt (gesehen von
beliebigen anderen Beobachtern) befindet. Ein solcher Beobachter durchlebt also wieder und wieder seine Historie --- die
ebenso seine Zukunftsgeschichte ist.  

Man kann argumentieren, dass zu dem Begriff einer Zeitmaschine/Zeitreisen mehr geh\"ort als das Auftreten von geschlossenen zeitartigen
Kurven. Insbesondere unterscheidet das Auftreten von geschlossenen zeitartigen Kurven nicht zwischen ``in die Zukunft reisen''
und ``in die Vergangenheit reisen'', und der Aspekt der (steuerbaren) ``Inbetriebnahme'' einer Zeitmaschine ist ebenso nicht
vertreten.
Um letzteren Punkt zu fassen, hat Hawking vorgeschlagen, dass Raumzeiten mit Zeitreisen so beschaffen sein sollen,
dass sie bis zu einem bestimmten Zeitpunkt eine v\"ollig regul\"are Raumzeitgeometrie aufweisen, in der keine
geschlossenen zeitartigen Kurven m\"oglich sind, und geschlossene zeitartige Kurven erst nach diesem Zeitpunkt
auftreten. Genauer bedeutet dies, dass geschlossene zeitartige Kurven nur mit einem ``kompakt generierten 
Cauchy-Horizont'' auftreten k\"onnen \cite{Haw92}. Der Cauchy-Horizont ist die Grenze, die den regul\"aren Bereich der
Raumzeit-Geometrie von dem mit geschlossenen zeitartigen Kurven trennt. Dieser soll kompakt generiert, also
gewissermassen endlich ausgedehnt sein, so dass auch die geschlossenen zeitartigen Kurven auf einen endlich ausgedehnten
Bereich in der Raumzeit beschr\"ankt sind. Das Ergebnis von Hawking besagt, dass ein Auftreten von geschlossenen zeitartigen Kurven
nur m\"oglich ist, wenn die Bedingung der positiven Energiedichte zumindest an einigen Punkten der Raumzeit verletzt wird,
was f\"ur die \"ublichen, makroskopischen Modellierungen von Energie-Materie-Verteilungen in der Regel ausgeschlossen ist.
An dieser Stelle mag darauf hingewiesen sein, dass auch die von Hawking gegebene Definition von Raumzeiten mit
Zeitreisen nicht alle Aspekte ber\"ucksichtigt, die mit Zeitreisen assoziiert werden k\"onnen.
So ist dabei der problematische Punkt von Zeitreisen in die Vergangenheit, mittels derer man die bereits faktische
Historie ab\"andern und auf diese Weise Paradoxien produzieren k\"onnte, nicht wirklich erfasst. F\"ur \"Uberlegungen dazu
sei auf die Literatur \cite{Nahin,EarSmeWut} und dort zitierte Referenzen verwiesen. 
Als eine weitere Bemerkung sei noch erw\"ahnt, dass die G\"odelsche Raumzeit \cite{Visser}
geschlossene zeitartige Kurven besitzt, aber auch eine positive Energiedichte (wenn man das Auftreten einer
von null verschiedenen kosmologischen Konstante akzeptiert), woraus sich ergibt, dass es keinen kompakt generierten
Cauchy-Horizont in der G\"odelschen Raumzeit gibt und damit kein wirkliches Zeitreise-Szenario.

Es gibt noch andere Raumzeiten, die Zeitreise-Szenarien zulassen. Prominente Beispiele sind Raumzeiten mit ``Wurml\"ochern''
oder Raumzeiten mit ``warp-drive'' Metriken. Raumzeiten mit Wurml\"ochern \cite{Visser} weisen ``Tunnel'' auf, durch die zwei
r\"aumlich weit voneinander entfernte Punkte wie durch eine Abk\"urzung in k\"urzerer Zeit verbunden werden k\"onnen
als durch den ``normalen'' Weg in der Raumzeit ohne Tunnel (siehe \cite{FordRoman-SciAm} f\"ur Illustration).
Kip Thorne hat gezeigt, dass eine bestimmte Form der
Aneinanderreihung solcher Wurml\"ocher oder Tunnel f\"ur Zeitreisen genutzt werden kann \cite{Visser}; dies ist verwandt
einem Argument, wonach Reisen mit \"Uberlichtgeschwindigkeit Zeitreisen erm\"oglichen \cite{Nahin}. 
Interessant w\"are die M\"oglichkeit, solche Tunnel oder Wurml\"ocher durch einen physikalischen
Proze{\ss} gezielt herzustellen, aber dies w\"urde bedeuten, eine Ver\"anderung in der Topologie der Raumzeit-Mannigfaltigkeit
herbeizuf\"uhren, jedoch scheint es au{\ss}erhalb einer hypothetischen Theorie der Quantengravitation bisher keinen
Hinweis darauf zu geben, was ein solcher physikalischer Prozess sein k\"onnte. Aber auch ohne die Frage anzugehen, wie
Wurml\"ocher in einer Raumzeit hergestellt werden k\"onnten, w\"are ihre blo{\ss}e Existenz fragw\"urdig, weil
es erheblicher negativer Energiedichten bed\"urfte, um sie f\"ur eine hinreichend lange Zeit und in gen\"ugender
Gr\"o{\ss}e aufrecht zu erhalten, so dass ein makroskopisches Objekt (oder ein Mensch) sie passieren kann (man spricht dann von
``durchschreitbaren Wurml\"ochern'') \cite{Visser}. Insofern wird eine Raumzeit mit (durchschreitbaren) Wurml\"ochern als
eine ``Designer-Raumzeit'' angesehen, d.h.\ als eine Raumzeit, deren Topologie und Metrik so konstruiert werden,
dass sie Zeitreisen erm\"oglichen, wobei aber in Kauf genommen wird, dass die linke Seite der Einsteinschen Feldgleichungen
unvertr\"aglich ist mit der Bedingung positiver Energiedichte auf der rechten Seite, d.h.\ dass es keine
bekannte (makroskopische) Form von Materie gibt, mit der sich eine solche Raumzeit-Geometrie konsistent mit den
Einsteinschen Feldgleichungen realisieren lie{\ss}e. In die Klasse solcher Designer-Raumzeiten fallen auch Raumzeiten
mit ``warp-drive'' Metriken \cite{Alcu,PfenRoman-warp}. Die Idee einer warp-drive Metrik ist es, auf einer Raumzeit mit einer vorgegebenen
Hintergrund-Metrik eine St\"orung der Metrik propagieren zu lassen, so dass ein Raumzeit-Gebiet innerhalb der
propagierenden Metrik-St\"orung eine h\"ohere Lichtgeschwindigkeit zul\"asst als die Hintergrund-Metrik. Die
Metrik St\"orung bezeichnet man auch als eine Blase (``bubble''), innerhalb derer die Lichtgeschwindigkeit durch
eine ge\"anderte Metrik gr\"o{\ss}er ist als au{\ss}erhalb. Zudem kann sich die Blase mit prinzipiell beliebiger
Geschwindigkeit bewegen, ohne dass dies zu einer Verletzung des Prinzips f\"uhrt, dass materielle Objekte
sich nicht schneller als mit Lichtgeschwindigkeit bewegen k\"onnen. Man kann dann zeigen, dass Objekte, die
auf eine solche Blase treffen, davon ein St\"uck weit mitgef\"uhrt werden und dadurch eine Geschwindigkeit erreichen,
die in Bezug auf die Hintergrund-Metrik oberhalb der Lichtgeschwindigkeit liegt, im Prinzip ohne
Beschr\"ankung. Man kann das durch einen Vergleich aus der Schiffahrt illustrieren. F\"ur Schiffe gibt es in
Verdr\"angerfahrt auf dem Wasser eine oberste Grenzgeschwindigkeit, die nicht \"uberschritten werden kann.
Es ist aber m\"oglich, dass ein Schiff auf eine Welle trifft, die sich sehr schnell ausbreitet und von
der das Schiff ein St\"uck weit mitgetragen wird, d.h.\ das Schiff ``surft'' auf der Welle. Auf diese Weise
kann das Schiff gegen\"uber dem ruhenden Wasser eine Fortbewegungsgeschwindigkeit erreichen, die gr\"o{\ss}er
ist als seine Grenzgeschwindigkeit in Verdr\"angerfahrt. Allerdings kann gezeigt werden, dass eine warp-drive
Metrik negative Energiedichten besitzen muss, und zwar an den R\"andern der Blase, also den Grenzen der
Metrik-St\"orung (obgleich diese so konstruiert ist, dass sie glatt in die Hintergrund-Metrik \"ubergeht, der
Rand der Blase ist also tats\"achlich gegl\"attet, und das Auftreten der negativen Energiedichten nicht etwa eine
Folge eines scharfen oder abrupten Randes der Blase). Auch hier ist der Vergleich mit dem Schiff, das auf
einer Welle surft, wieder suggestiv: Der Wellenberg (entsprechend positiver Energiedichte), auf dem das
Schiff surft, ben\"otigt auch benachbarte Wellent\"aler (entsprechend negativer Energiedichte).
Also ist auch hier wieder fraglich, ob warp-drive Metriken realistisch sind; in jedem Fall f\"uhrten sie zu
eigenartigen Effekten f\"ur Beobachter, die auf die Blasen treffen, wor\"uber an anderer Stelle berichtet
wurde \cite{Tippet}.

Alle bisher vorgestellten Designer-Raumzeiten scheinen nicht als L\"osungen der Einsteinschen
Feldgleichungen mit bekannten Formen makroskopischer Energie-Materie-Verteil\-un\-gen realisierbar zu sein,
da sie eine Verletzung der Positivit\"at der Energiedichte erfordern, d.h.\ f\"ur solche Raumzeit-Geometrien
gibt es Punkte $x$ in der Raumzeit $M$, f\"ur die 
$$ \varrho[u](x) = T_{ab}(x)u^a u^b < 0 \quad \ \ (u^a = u^a(x))$$
gilt. Das bedeutet, dass an einigen Orten zu bestimmten Zeiten eine negative Energiedichte auftreten muss.
W\"ahrend das, wie schon betont, f\"ur makroskopisch beschriebene Materie \"ublicherweise nicht eintritt,
besteht eine solche M\"oglichkeit in der Quantenphysik. Hier kommt also endlich die Quantenphysik ins Spiel,
deren Besonderheiten im Vergleich zur makroskopischen Beschreibung von Materie zu ganz neuen Situationen
f\"uhren k\"onnten.

\section{Quantenfeldtheorie (und Gravitation), semiklassische Einstein-Gleichung}

Grob gesprochen besteht ein Grundgedanke der Quantenfeldtheorie darin, die (in jeder Vektor- oder Tensorkomponente) zahlenwertigen
Felder, die die Freiheitsgrade der Materie beschreiben, durch operatorwertige Felder zu ersetzen. Alle solche
Felder, die observablen Gr\"o{\ss}en entsprechen, bilden dann eine nicht-kommutative Algebra, die zus\"atzlich
Lokalit\"ats- , Kovarianz- und Kausalit\"atsstrukturen tr\"agt (das soll an dieser Stelle
nicht weiter ausgef\"uhrt werden, siehe dazu die Referenzen \cite{BFV,WaldQFT,BDH}). Dar\"uber hinaus wird eine
Menge von Zust\"anden ben\"otigt, in denen Erwartungswerte der Observablen gebildet werden k\"onnen. Diese Zustandsmenge
soll so gew\"ahlt sein, dass Zust\"ande darin enthalten sind, die dem Bild einer Situation mit einer niedrigen
Teilchen- und Energiedichte und einer m\"oglichst geringen Abweichung von einer Gleichgewichtssituation entsprechen.
F\"ur Quantenfelder auf der Minkowski-Raumzeit (oder auf Raumzeiten, die eine gen\"ugend reichhaltige Isometriegruppe
besitzen) gibt es f\"ur diese Zustandsmenge eine kanonische Wahl, da man von einem Vakuum-Zustand ausgehen kann,
der sich durch sein Verhalten gegen\"uber der Wirkung der Isometrien der Raumzeit auszeichnen l\"asst \cite{Haag}. Im allgemeinen
besitzen Raumzeiten mit beliebigen Metriken aber keine gen\"ugend reichhaltigen Symmetriegruppen, die es gestatten
w\"urden, einen Vakuumzustand auszuzeichnen. Dies hat auch zur Folge, dass der Teilchenbegriff beobachterabh\"angig wird.
Das wird beispielsweise deutlich beim Hawking-Effekt und beim Unruh-Effekt \cite{Haw75,Unruh,Haag,BuSolveen}. 
Das Kriterium, das nunmehr zur
Charakterisierung einer Zustandmenge nach den genannten Ma{\ss}gaben als akzeptiert gilt, wird mathematisch so formuliert:
Die Zust\"ande sollen die ``mikrolokale Spektrumsbedingung'' erf\"ullen. Auf die genaue Definition werden wir nicht
eingehen, da sie einen gewissen mathematischen Aufwand erfordert. F\"ur Details sei auf die Referenzen \cite{BFK,SVW,BDH} verwiesen.
Wir bemerken nur, dass die mikrolokale Spektrumsbedingung f\"ur lineare Quantenfelder (wozu unter anderem das Klein-Gordon-Feld,
das Dirac-Feld und das freie elektromagnetische Feld geh\"oren) \"aquivalent ist zur Hadamard-Bedingung, die wichtig ist um
die Erwartungswerte des quantisierten Energie-Impuls-Tensors systematisch zu definieren, was auch eine Motivation f\"ur
diese Bedingung gewesen ist \cite{WaldQFT}. Ferner gibt es Beziehungen zwischen der 
mikrolokalen Spektrumsbedingung und Bedingungen der thermodynamischen Stabilit\"at, die im wesentlichen die G\"ultigkeit
des 2.\ Hauptsatzes der Thermodynamik aussagen \cite{FewVer-Stability}. 

 Man kann also davon ausgehen, dass sich f\"ur quantisierte Felder in Zust\"anden $\psi$, die
die  ``mikrolokale Spektrumsbedingung'' erf\"ullen, die Erwartungswerte $\langle {\sf T}_{ab}(x) \rangle_{\psi}$
des quantisierten Energie-Impuls-Tensors (symbolisch mit ${\sf T}_{ab}(x)$ bezeichnet) an jedem Punkt $x$ der Raumzeit definieren lassen.
Tats\"achlich ist die Definition nicht unbedingt eindeutig; f\"ur eine Diskussion dazu siehe \cite{Ver-Reb}.   Im folgenden k\"onnen wir davon
ausgehen, dass eine eindeutige Definition spezifiziert worden ist. Die weiter unten erw\"ahnten Ergebnisse h\"angen nicht davon ab, in welcher Weise das 
erreicht wurde. 

Unter sehr allgemeinen Bedingungen kann dann gezeigt werden, dass die Erwartungswerte
$\langle {\sf T}_{ab}(x) \rangle_{\psi}$ die
Bedingung der Positivit\"at der Energiedichte verletzen k\"onnen (und in gewisser Weise dies sogar tun m\"ussen).
F\"ur lineare Quantenfelder ist sogar folgendes bekannt: Zu einem gegebenen Raumzeit-Punkt $x$ und zeitartigem Vektor $u^a$ bei $x$ gibt
es eine Folge von Zust\"anden $\psi_n$, $n \in \mathbb{N}$, so dass die Energiedichte-Erwartungswerte in dieser Zustandsfolge immer
st\"arker negativ werden \cite{Few-LectN}, d.h. 
$$  \langle {\sf T}_{ab}(x) \rangle_{\psi_n}u^a u^b \to - 
\infty \quad \text{f\"ur} \quad n \to \infty\,. $$ 
Diese Eigenschaft der Erwartungswerte des Energie-Impuls-Tensors kann durchaus Folgen haben,
wenn man diese Erwartungswerte auf die rechte Seite der Einsteinschen Feldgleichungen schreibt:
$$ R_{ab} - \frac{1}{2}g_{ab}R + \Lambda g_{ab} = 
\langle {\sf T}_{ab} \rangle_{\psi}
 $$
Diese Gleichung wird als semiklassische Einstein-Gleichung bezeichnet; die Raumzeit-Geo\-metrie
wird im Sinne einer klassischen, d.h.\ nicht quantisierten Feldtheorie behandelt, die 
in der Raumzeit vorhandene Materie (oder Strahlung) aber im Sinne der Quantenfeldtheorie.
F\"ur ein gegebenes lokal kovariantes Quantenfeld ist eine L\"osung der semiklassischen
Einstein-Gleichung gegeben durch eine Metrik $g_{ab}$ und einen Zustand $\psi$ des 
Quantenfelds so, dass die Gleichung erf\"ullt ist (ggf.\ zu gegebenen Anfangsbedingungen).
Dabei sollten die Varianzen des Energie-Impuls-Tensors im Zustand $\psi$ nicht zu gro{\ss}
sein, weil dies die semiklassische Approximation fraglich machen w\"urde. In der letzten
Zeit wurden einige Fortschritte im Zusammenhang mit der semiklassischen Einstein-Gleichung
erzielt, siehe z.B. \cite{EltGot,Pina}. 

\section{Zeitreisen mit Quantenfeldtheorie-Materie?}

Die Tatsache, dass es m\"oglich ist, an gegebenen Raumzeit-Punkten $x$ die Positivit\"at
des Erwartungswerts der Energiedichte beliebig zu verletzen, scheint die M\"oglichkeit
zu er\"offnen, Designer-Raumzeiten, insbesondere solche, die Zeitreisen zulassen, als
L\"osungen der semiklassischen Einstein-Gleichung und damit als physikalisch realisierbare
Situationen erhalten zu k\"onnen, sofern die entsprechenden Quantenfeld-Zust\"ande
hergestellt werden k\"onnen. Das w\"aren spannende Aussichten auf der Schnittfl\"ache von
Gravitation und Quantenphysik! 

Es zeigt sich jedoch, dass die Aussichten so spektakul\"ar nicht sind, wie es zun\"achst den
Anschein hat. Ein erstes Ergebnis von Kay, Radzikowski und Wald \cite{KRW} zeigt die Unvertr\"aglichkeit von Zeitreisen-Raumzeiten mit
der Wohldefiniertheit der Energie-Impuls-Tensor Erwartungswerte und der mikrolokalen Spektrumsbedingung.
Wir geben es hier in der folgenden Form wieder (f\"ur die pr\"azise Aussage siehe die Originalarbeit \cite{KRW}):
\\[6pt]
{\it Die Erwartungswerte des Energie-Impuls-Tensors $\langle {\sf T}_{ab}(x) \rangle_{\psi}$ m\"ussen f\"ur Zust\"ande
$\psi$, die im regul\"aren Teil einer Raumzeit mit kompakt generierten Cauchy-Horizont die mikrolokale Spektrumsbedingung
erf\"ullen, an den Punkten $x$ des Cauchy-Horizonts singul\"ar werden (bzw.\ die Erwartungswerte sind an diesen
Punkten nicht konsistent definierbar).}
\\[6pt]
Mit anderen Worten: Die Eigenschaften der Energie- und Impulsdichten von Quantenfeldern m\"ussen an der ``Trennfl\"ache''
zwischen dem regul\"aren Teil der Raumzeit, in dem keine Zeitreisen vorkommen k\"onnen, und dem Teil, in dem
Zeitreisen ``durch Bet\"atigung einer Zeitmaschine in Gang gesetzt werden'', pathologisch werden --- sofern f\"ur
ihre Beschreibung dieselben Regeln angewendet werden wie im regul\"aren Teil der Raumzeit. Dies kann als ein ziemlich
starkes Argument gegen die M\"oglichkeit von Zeitreisen unter Verwendung von quantisiert beschriebener Materie angesehen werden.
\\[6pt]
Weitere Argumente gegen Designer-Raumzeiten lassen sich daraus ableiten, dass zwar an einem fest vorgegebenen Punkt $x$ der
Raumzeit der Erwartunswert der Energiedichte $\langle {\sf T}_{ab}(x) \rangle_{\psi}u^au^b$ durch Wahl eines geeigneten Zustands
$\psi$ beliebig negativ gemacht werden kann, dass dies aber nicht mehr der Fall ist, wenn man statt der Energiedichte an einem
festen Punkt eine zeitliche Mittelung der Energiedichte betrachtet wird, die ein Beobachter oder ein Messger\"at entlang einer
Weltlinie registrieren. Genauer gesagt gilt folgendes. Betrachtet wird ein lineares quantisiertes Feld auf
einer Raumzeit $M$ mit Metrik $g_{ab}$. Es sei $\gamma: I \to M, t \mapsto \gamma(t)$ eine zukunftsgerichtete,
zeitartige Kurve mit Tangente $u^a = (d \gamma/dt)^a$ entlang der Kurve. Ferner sei $g: I \to \mathbb{R}$, $t \mapsto g(t)$
eine glatte Funktion, die au{\ss}erhalb eines festen, echten Teilintervalls von $I$ gleich $0$ ist. Dann gibt es eine
Konstante Zahl $q$ so, dass die Absch\"atzung
$$ \min_{\psi}\, \int dt\, (g(t))^2\, \langle {\sf T}_{ab}(\gamma(t)) \rangle_{\psi}u^au^b \ge q $$
gibt. Das Integral mit der glatten, nicht-negativen Gewichtsfunktion $g^2$ ist ein gewichtetes zeitliches Mittel der
von dem Beobachter mit Weltlinie $\gamma$ registrierten Energiedichte im Zustand $\psi$ des Quantenfelds auf der Raumzeit.
Zu beachten ist, dass auf der linken Seite der Ungleichung das Minimum \"uber alle Zust\"ande $\psi$, die die mikrolokale Spektrumsbedingung
erf\"ullen, gebildet wird, und dass die endliche Zahl $q$ auf der rechten Seite (die m\"oglicherweise negativ sein kann), nicht von
$\psi$ abh\"angt. Tats\"achlich h\"angt $q$ (nur) von der Weltlinien-Kurve $\gamma$ und von der Wahl der Gewichtsfunktion $g$ ab,
$q = q(\gamma,g)$. Es ist also bei fester Wahl von $\gamma$ und $g$ nicht m\"oglich, durch Wahl einer geeigneten Folge von
Zust\"anden die zeitlich gemittelte Energiedichte beliebig negativ zu machen, so wie das m\"oglich ist, wenn die Energiedichte
an einem gew\"ahlten Punkt betrachtet wird. Solche Ungleichungen werden als ``Quanten-Energie-Ungleichungen'' bezeichnet. Die G\"ultigkeit
solcher Ungleichungen wurden zuerst von L.\ Ford vorgeschlagen und in speziellen F\"allen untersucht \cite{Ford}; die erste allgemeine Quanten-Energie-Ungleichung
f\"ur lineare skalare Quantenfelder auf generischen Raumzeiten wurde von C.J. Fewster \cite{Few-QWEI1} etabliert. Es hat sich gezeigt, dass 
Quanten-Energie-Ungleichungen f\"ur L\"osungen der semiklassischen Einstein-Gleichungen erhebliche Einschr\"ankungen an die M\"oglichkeit von
Wurmloch- und warp-drive-Raumzeiten ergeben, sowie f\"ur einige andere Szenarien. 
F\"ur eine ausf\"uhrliche Darstellung sei auf eine Vorlesungsausarbeitung von
Fewster verwiesen \cite{Few-LectN}.
Quanten-Energie-Ungleichungen sagen aus, dass die Energiedichte zwar beliebig negativ gemacht werden kann, aber
dass dies dann nur f\"ur eine kurze Zeitspanne aufrecht erhalten werden kann. Dies f\"uhrt zu Einschr\"ankungen,
wieviel negative Energie f\"ur wie lange oder wie weitr\"aumig verteilt angeh\"auft werden kann. Solche
Situationen
wurden vor allem von L.\ Ford, T.\ Roman und Koautoren untersucht. Wir geben ihre wichtigsten Ergebnisse hier
wie folgt wieder --- f\"ur eine wesentlich genauere Diskussion sei verwiesen auf
\cite{FordRoman-wurm}, \cite{PfenRoman-warp}, \cite{FordRoman-SciAm} und \cite{Roman}.
\\[6pt]
{\it F\"ur die Aufrechterhaltung eines durchschreitbaren Wurmlochs mit einem Durchmesser von einem Meter ist
es erforderlich, einen negativen Energieinhalt, der der Strahlungsleistung von $10^9$ Sternen in einem Jahr
entspricht, in einem Volumen von einem Durchmesser von ca.\ $10^{-21}$ Meter (dem Millionstel des Protonradius)
anzusammeln.}
\\[6pt]
{\it Eine warp-drive Blase, mit der ein Raumschiff von 200 Metern Gr\"o{\ss}e mit 10-facher Lichtgeschwindigkeit
transportiert werden k\"onnte, m\"usste einen Blasenrand von nur $10^{-32}$ Metern Dicke haben, und es m\"usste im
Blasenrand ein negativer Energieinhalt angeh\"auft werden, der die Masse des sichtbaren Universums
um den Faktor $10^{10}$ \"ubersteigt.}
\\[6pt]
Eine solche Anh\"aufung von negativer Energie derart gigantischen Ausma{\ss}es in sub-mikros\-kopischen
Raumbereichen erscheint vollkommen unrealistisch. 

Man kann also festhalten, dass Zeitreisen-Szenarien 
f\"ur makroskopische Reisende als L\"osungen der semiklassischen Einstein-Gleichungen
nicht vorkommen k\"onnen. Nicht v\"ollig ausgeschlossen ist das Auftreten von sub-mikroskopischen
Wurml\"ochern unterhalb einer Gr\"o{\ss}e von etwa $10^{-32}$ Metern \cite{FordRoman-wurm}, was schon sehr nahe
an der G\"ultigkeitsgrenze der semiklassischen Einstein-Gleichung liegt (in diesem
Gr\"o{\ss}enordnungsbereich w\"are die Behandlung der Gravitation als quantisierte Theorie
erforderlich). Insgesamt ist das wenig verwunderlich: Die Effekte, in denen die Quantenphysik
deutlich von der makroskopischen Physik abweicht, zeigen sich typischerweise auf sehr kleinen
Zeit- und L\"angenskalen; makroskopische Zeitreise-Szenarien liegen nahe an der makroskopischen
Modellierung von Energie und Materie, f\"ur die die Positivit\"at der Energiedichte typischerweise
erf\"ullt ist, was wiederum das Auftreten von Zeitmaschinen und Zeitreisen weitgehend ausschliesst.
Im Sinne von Hawking \cite{Haw92} l\"asst sich konstatieren, dass das Zusammenspiel von Gravitation
und Quantenphysik sehr sicher ist gegen\"uber Versuchen, die faktische Historie von allem, was sich
in Raum und Zeit ereignet hat, mit Hilfe von Zeitreisen(den) abzu\"andern.
${}$\\[20pt]
{\bf \large Danksagung}\\[6pt]
Der Autor dankt C.\ Dappiaggi, C.J.\ Fewster, L.H.\ Ford, T.-P.\ Hack und T.A.\ Roman f\"ur
wertvolle Hinweise.

\end{document}